\documentstyle[prl,aps,epsfig,twocolumn]{revtex}

\begin{document}
\draft
\title{New Universality Class at the Magnetic Field Tuned Superconductor-Insulator
Transition?}
\author{N. Markovi\'{c}, C. Christiansen and A. M. Goldman}
\address{School of Physics and Astronomy, University of Minnesota, Minneapolis,\\
MN 55455, USA}
\date{June 1, 1998}
\maketitle

\begin{abstract}
The superconductor-insulator transition in ultrathin films of amorphous Bi
was tuned by an applied magnetic field. A finite size scaling analysis has
been carried out to determine the critical exponent product $\nu z=$ $0.7\pm
0.1$ for five films of different thicknesses and normal state resistances.
This result differs from the exponents found in previous experiments on InO$%
_{x}$ and MoGe films. We discuss several possible reasons for this
disagreement.
\end{abstract}

\pacs{PACS numbers: 74.76.-w, 74.40.+k, 72.15.Rn}

The superconductor-insulator (SI) transition in two-dimensional disordered
systems has been extensively studied theoretically in the context of the
dirty boson model at zero temperature. Monte Carlo simulations \cite{Monte
Carlo,Cha} and renormalization group calculations \cite{RG} mostly focused
on the case when the transition is tuned by changing the carrier
concentration, coupling strength or the strength of the disorder. The SI
transition can also be tuned by changing the external magnetic field \cite
{Fisher B}. Scaling arguments \cite{Sondhi} give a lower bound on the
correlation length exponent $\nu \geq 1,$ and predict the value of the
dynamical critical exponent to be z=1 \cite{Fisher B,Fisher g}. An
experimental scaling analysis of the magnetic field tuned SI transition has
been carried out for thin films of $In0_{x}$ \cite{Hebard} and MoGe \cite
{Yazdani}. Both experiments found $\nu z\approx 1.3,$ consistent with the
theoretical predictions. We have studied the magnetic field tuned SI
transition in amorphous ultrathin homogeneous films of Bi. Finite size
scaling analysis yields $\nu z\approx 0.7$ which is in disagreement with the
predictions of Ref. \cite{Fisher B}, as well as with the results of Refs. 
\cite{Hebard} and \cite{Yazdani}.

The investigations were carried out on five ultrathin Bi films with
thicknesses close to 12-13\AA . The films were evaporated on top of a 10\AA\
thick layer of amorphous Ge, which was pre-deposited onto a 0.75mm thick
single-crystal of $SrTiO_{3}$ $(100)$. The film thickness was systematically
increased $in$ $situ$ under UHV conditions ($\sim 10^{-10}$Torr), with the
substrate temperature kept below 20 K during all depositions. Film
thicknesses were determined using a previously calibrated quartz crystal
monitor. Films prepared in this manner are believed to be homogeneous, since
it has been found that they become connected at an average thickness on the
order of one monolayer \cite{Strongin}. Between the successive depositions,
resistance measurements were carried out using a standard dc four-probe
technique with current bias up to 50nA. Magnetic fields up to 12kG
perpendicular to the plane of the sample were applied using a
superconducting split-coil magnet.

A typical temperature dependence of the resistance as the magnetic field
changes is shown on Fig.\ref{R vs T for B}. In zero field, the resistance
decreases with increasing temperature suggesting the existence of
superconducting fluctuations. Magnetic field destroys this downward
curvature, and at some critical magnetic field, $B_{c},$ the resistance is
independent of temperature. In magnetic fields higher than B$_{c}$ the film
is insulating, with $\delta R/\delta T>0$. The critical magnetic field is
determined by plotting the resistance as a function of magnetic field for
different temperatures (Inset of Fig.\ref{R vs T for B}) and identifying the
crossing point for which the resistance is temperature independent, or by
plotting $dR/dT$ as a function of magnetic field at the lowest temperatures
and finding the field for which $(dR/dT)=0$ \cite{Hebard}.

The scaling relation for the resistance of a two dimensional system close to
the magnetic field tuned superconductor insulator transition was suggested
by Fisher \cite{Fisher B}:

\begin{equation}
R(\delta ,T)=R_{c}\ f(\delta T^{-1/\nu z})  \label{one}
\end{equation}
Here $\delta =B-B_{c}$ is the deviation from the critical field, $R_{c}$ is
the critical resistance at $B=B_{c}$, $f(x)$ is a universal scaling function
such that $f(0)=1$, $\nu $ is the coherence length exponent, and $z$ is the
dynamical critical exponent. To carry out the analysis of the data, we
introduce a parameter $t\equiv T^{-1/\nu z},$ and adjust $t(T)$ at each
temperature so that all the resistance data collapse on a single curve. The
critical exponent product $\nu z$ is then found from the temperature
dependence of $t$. This procedure does not require prior knowledge of the
critical exponents or detailed knowledge of the functional form of the
temperature or field dependence of the resistance.

The collapse of the resistance data as a function of $\delta t$ for one of
the samples is shown in Fig.\ref{B collapse}. The critical exponent product $%
\nu z$, determined from the temperature dependence of the parameter t (Inset
of Fig.\ref{B collapse}), is found to be $\nu z=0.7\pm 0.1.$ The same
exponents were obtained using a somewhat different method of plotting $
(dR/dB)|_{B_{c}}$ vs $T^{-1}$ on a log-log plot and determining the slope
which is equal to $1/\nu z$ \cite{Hebard}. The analysis was carried out for
five films of different thicknesses, all yielding $\nu z=0.7\pm 0.1$,
apparently independent of the film thickness. In fact, plotting the
normalized resistance $R/R_{c}$ as a function of the scaling variable
reveals that they all scale together (Fig. \ref{R/Rc collapse}). This
behavior strongly suggests that the scaling function is universal.

The critical resistance, however, does not seem to be universal. Figure \ref
{Rc vs B} shows that R$_{c}$ decreases as the critical field increases,
roughly in a linear fashion. Very similar behavior was observed by Yazdani
and Kapitulnik \cite{Yazdani}.

The critical exponent product $\nu z$ found in this experiment does not
agree with the theoretical prediction that $\nu \geq 1$ and $z=1$ \cite
{Fisher B} for a disordered system, nor with the experimentally obtained $%
z=1 $ and $\nu \approx 1.3$ for a magnetic field tuned SI transition in thin
films of amorphous $InO_{x}$ \cite{Hebard} and MoGe \cite{Yazdani}. There
could be a number of possible reasons for this disagreement, which we
discuss next.

The prediction that $\nu \geq 1$ follows from an exact theorem \cite{Chayes}
which is expected to hold for any transition which can be tuned by changing
the strength of the disorder. It has been argued recently \cite{Pazmandi}
that this is not necessarily the case, and that the disorder averaging may
play an important role in determining the critical exponents. The nature of
disorder would be crucial here.

Without disorder, the SI transition in a 2D system might belong to the
universality class of the classical 3D XY model\cite{Fisher B,Fisher g}. The
exponent product we find for the field-driven transition is very close to
that obtained for the classical 3D XY model. Numerical simulations of a T=0
critical point for the boson Hubbard model in 2D for which there is no
disorder \cite{Cha} also find $\nu z\approx 0.7.$ Since our films are
believed to be very homogeneous, it is possible that the length scale which
characterizes the disorder is very small compared to the diverging coherence
length, and the system behaves as if there were no disorder.

The importance of the length scale and the nature of disorder might also
explain the difference between our results and those of Refs. \cite{Hebard}
and \cite{Yazdani}. Since both $InO_{x}$ and MoGe are composite materials,
the length scale for disorder might be quite different from that of
monoatomic amorphous Bi. Indeed, $InO_{x}$ films in a magnetic field appear
to be in a quasi-reentrant regime, suggestive of the behavior of granular
films in zero field \cite{Liu}. Second, the way the samples are prepared
might result in a profoundly different nature of the disorder. In $InO_{x}$
and MoGe, the normal state resistance is determined by the composition
(content of oxygen and germanium, respectively), while in quench condensed
Bi films it depends on the film thickness. We also wish to point out that
the Bi films used in this study are almost an order of magnitude thinner
than the samples of Refs. \cite{Hebard} and \cite{Yazdani}.

It has been suggested that a local dissipation for the phase of the
superconducting order parameter due to gapless electronic excitations might
change the universality class of the system and lead to a non-universal
critical resistance \cite{Wagenblast}. The critical resistance is expected
to increase with increasing damping due to dissipation, which might be
expected to increase with decreasing normal state resistance. We however
observe that the critical resistance decreases as the normal state
resistance goes down, which is exactly the opposite of the behavior
predicted by Ref.\cite{Wagenblast}.

A recent work by Shimshoni et al. \cite{Shimshoni} considers the system
close to the SI transition to be a binary composite of an insulating and a
superconducting phase. Using incoherent Boltzman transport theory, they
derive resistivity laws in different temperature regimes and predict finite
dissipation at T=0 at all fields. We do not observe any saturation in the
temperature dependence of the resistance as the temperature decreases ($
\delta R/\delta T$ is non-zero down to the lowest temperatures). A
satisfactory fit to the predicted resistivity laws also could not be
obtained.

Local fluctuations of the amplitude of the superconducting order parameter
are usually neglected in the scaling theory and the numerical simulations,
but might actually be important. It has been suggested that the percolation
of islands with strong amplitude fluctuations might mask the localization
exponents obtained from the scaling theory \cite{Ramakrishnan}.

The existence of electronic excitations was also used to explain the
non-universal behavior of the critical resistance in Ref. \cite{Yazdani}. A
model of two-channel conduction was proposed, where the conductance due to
the electron channel adds to the conductance due to the boson channel. Films
with lower normal state resistances would have lower critical resistances
due to a larger fraction of normal electrons. Our data could only be
explained this way if the quantum resistance due to pairs were {\it greater}
than $h/4e^{2}$. The conductance due to the electronic channel in a magnetic
field might also depend on the strength of the spin-orbit interactions,
which is another major difference between our samples and those of Refs. 
\cite{Hebard} and \cite{Yazdani}. There is, however, a striking similarity
in the magnetic field and normal state resistance behavior of the critical
resistance.

We should note that the critical resistance is predicted to be universal
only at T=0, while the finite temperature corrections are expected to be
scaled by T$_{c}$ \cite{Fisher B}. Normal state resistances of the MoGe
films of Ref. \cite{Yazdani} are a factor of 3-10 lower than the Bi films
considered here, and are far away from the critical point of the disorder
driven transition. This means that our samples are probing a different part
of the phase diagram \cite{Fisher B,Markovic}, and the finite temperature
corrections might be more important in one case then the other. There is
also a question of whether any of the experiments really accessed the
quantum critical region \cite{Sachdev}.

In order to resolve some of the issues discussed in this paper, simulations
in which the magnetic field is the tuning parameter would be very useful.

We gratefully acknowledge useful discussions with A. P. Young, S. Sachdev
and S. L. Sondhi. This work was supported in part by the National Science
Foundation under Grant No. NSF/DMR-9623477.

\begin{figure}[tbp]
\caption{Resistance per square as a function of temperature in different
magnetic fields, ranging from 0kG (bottom) to 12kG (top), with 1kG
increments. Inset: Resistance per square as a function of magnetic field at
0.15, 0.17, 0.19, 0.2, 0.25, 0.3 and 0.35K.}
\label{R vs T for B}
\end{figure}

\begin{figure}[tbp]
\caption{Resistance per square as a function of the scaling variable, $%
t|B-B_{c}|$, for seventeen different temperatures, ranging from 0.14K to
0.5K. Here $t=T^{-1/\nu z}$ is treated as an adjustable parameter to obtain
the best collapse of the data. Different symbols represent different
temperatures. Inset: Fitting the temperature dependence of the parameter t
to a power law determines the value of $\nu z.$}
\label{B collapse}
\end{figure}

\begin{figure}[tbp]
\caption{Normalized resistance per square as a function of the scaling
variable, $T^{-1/\nu z}|B-B_{c}|$, at 0.15K for five films: 12.239\AA\
(squares), 12.353\AA\ (diamonds), 12.455\AA\ (crosses), 12.553\AA\
(triangles) and 12.660\AA\ (circles). Only one temperature was shown for
each film for clarity.}
\label{R/Rc collapse}
\end{figure}

\begin{figure}[tbp]
\caption{The critical resistance as a function of the critical field for the
same five films of Fig.3. Here R$_{c}$ decreases with increasing thickness,
as thicker films have lower normal state resistances and higher critical
fields.}
\label{Rc vs B}
\end{figure}

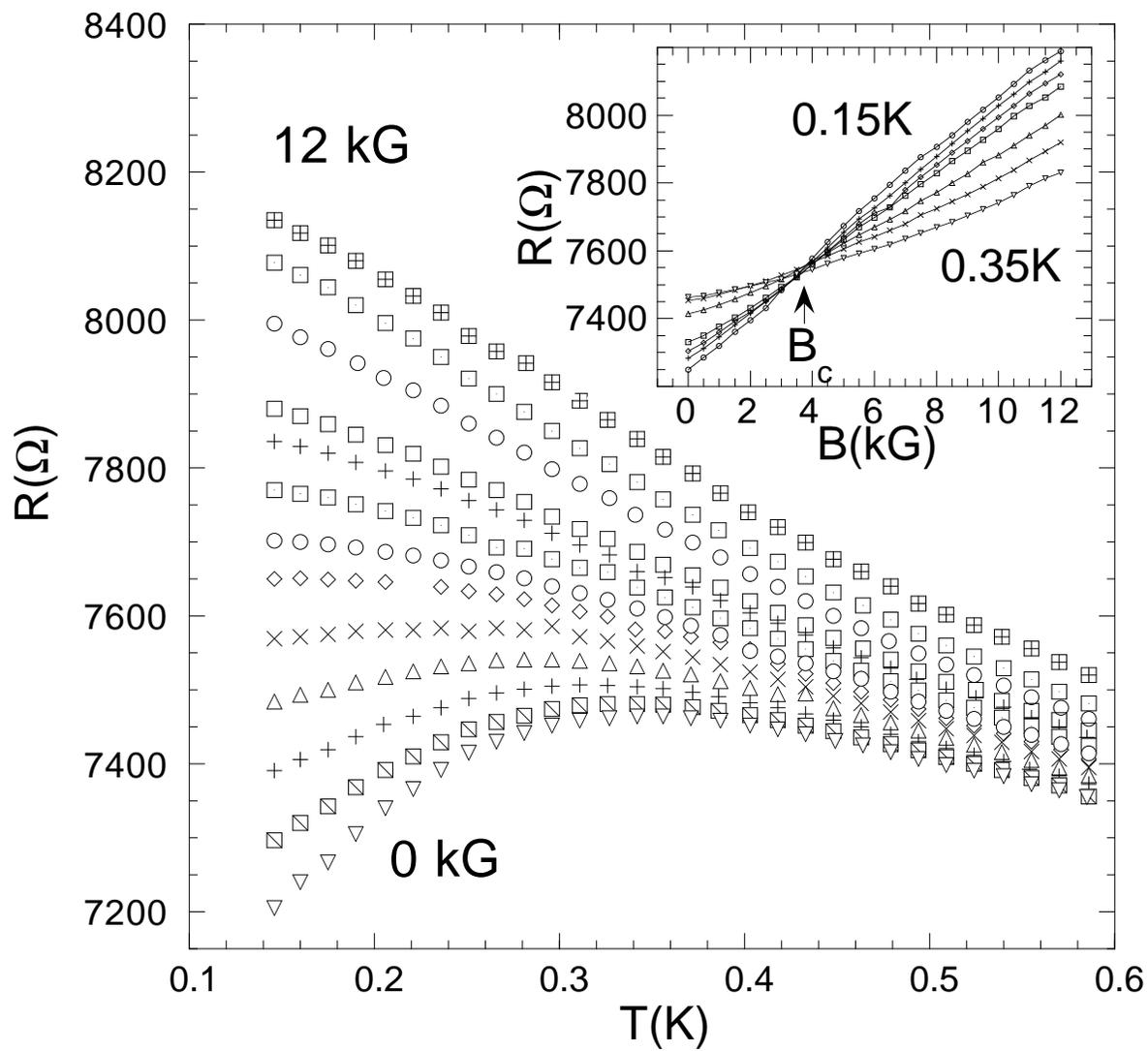
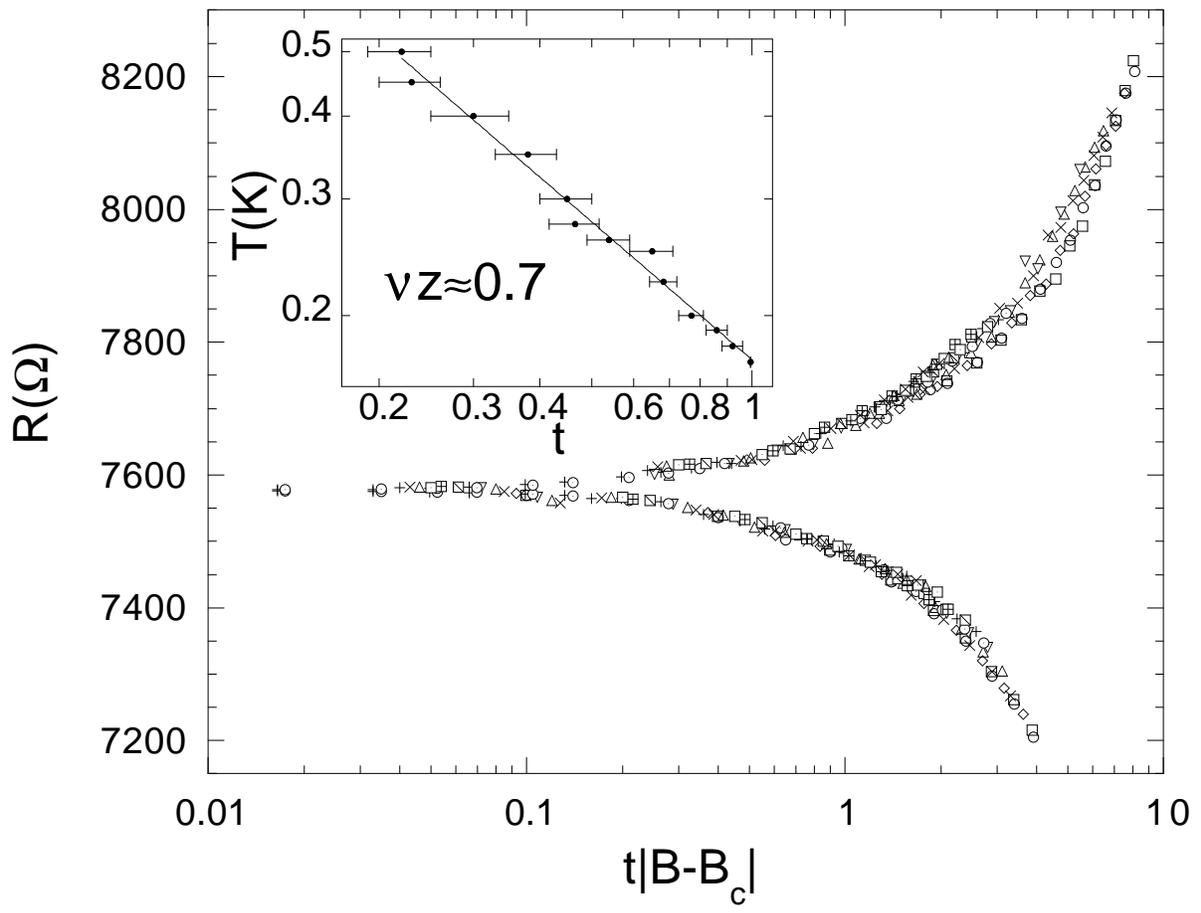
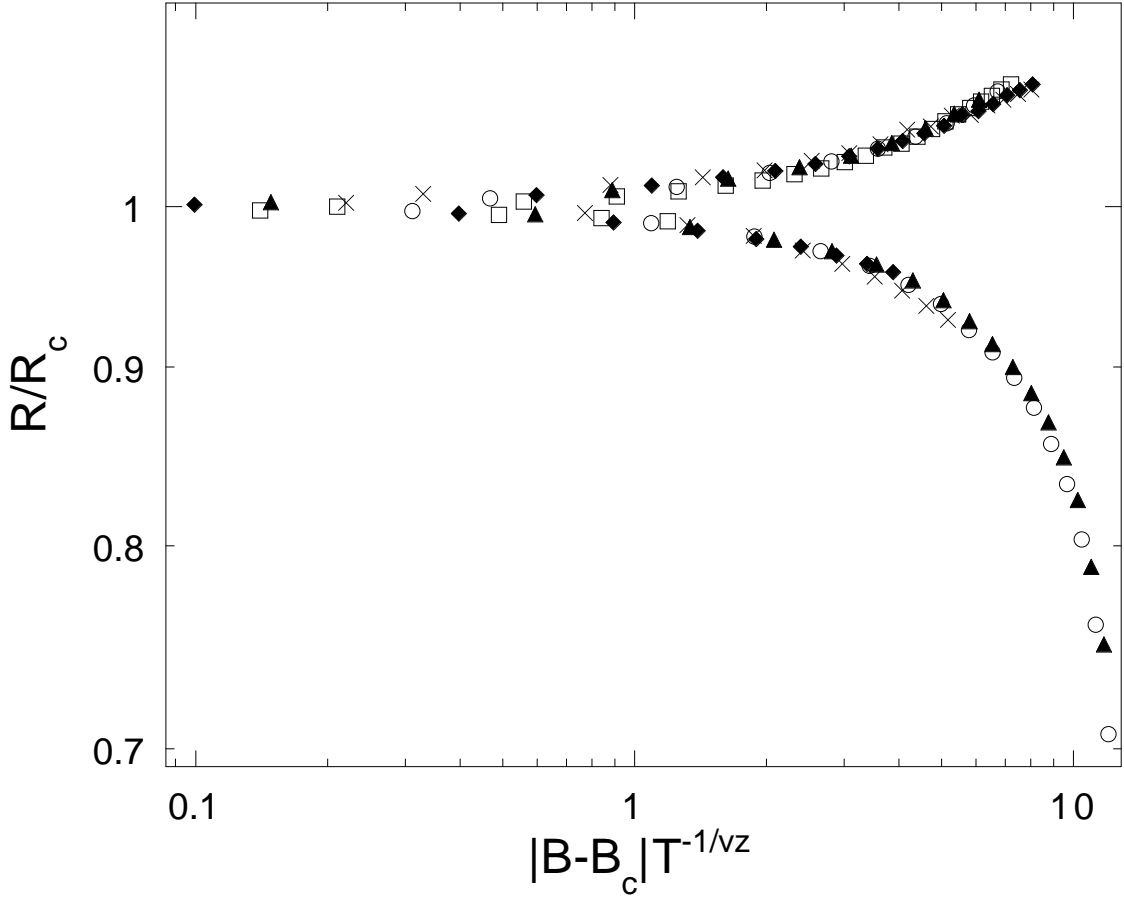
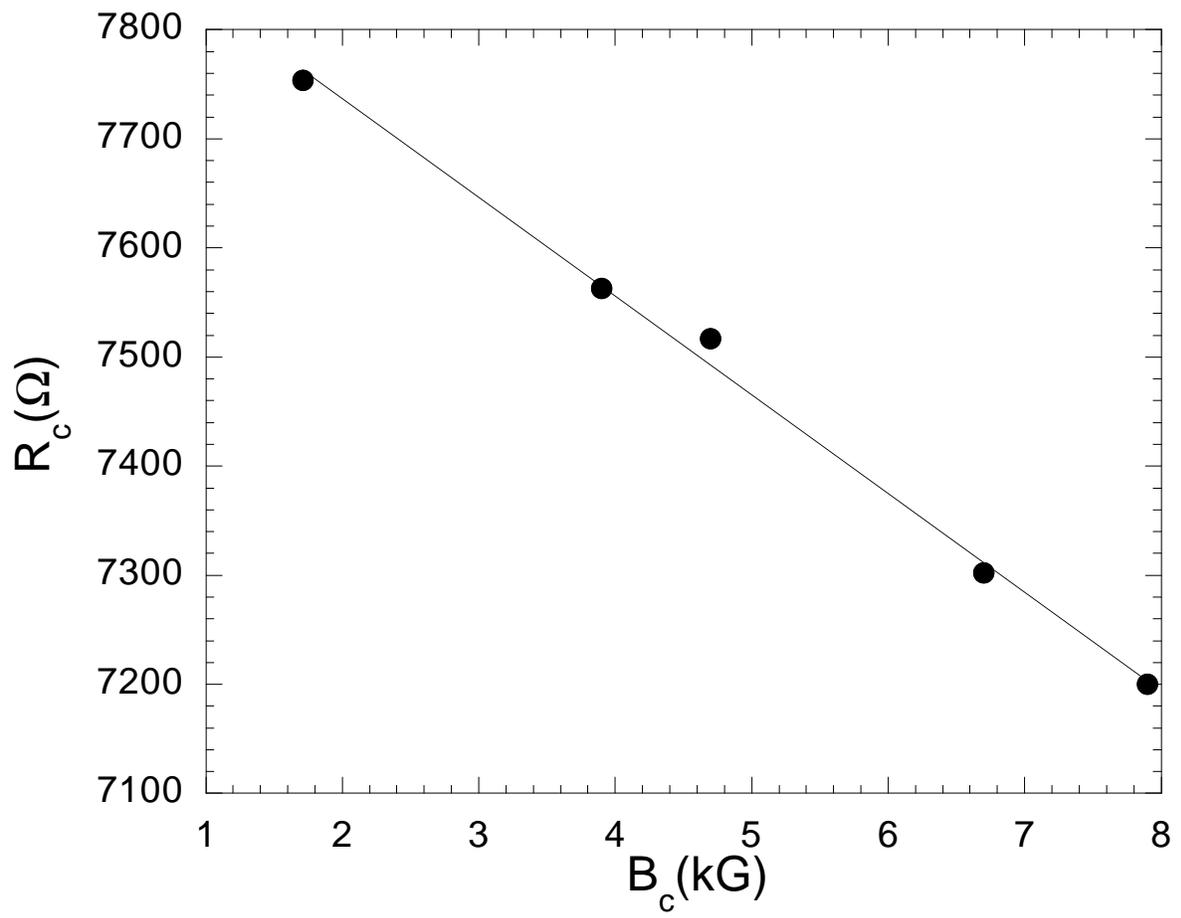
\end{document}